\RequirePackage{fix-cm}
\documentclass[smallcondensed]{svjour3ujm}     
\smartqed  
\usepackage{graphicx}
\usepackage[authoryear]{natbib}
\newcommand\bq{\par\begin{quote}}
\newcommand\eq{\par\end{quote}}

\journalname{}
\begin{document}
\title{The elision of the subject and the manifestation of the world}
\author{U.J. Mohrhoff}
\institute{Ulrich Mohrhoff \at
              Sri Aurobindo International Centre of Education\\
              605002 Pondicherry India\\
              \email{ujmohrhoff@gmail.com}}
\maketitle
\vspace{-36pt}
\begin{abstract}
Owing to the contextuality of the properties of quantum objects, quantum mechanics does not appear to countenance the elision of the thinking and perceiving subject. If quantum objects owe their properties to the experimental conditions in which they are observed, the experimental apparatus cannot owe its properties to the quantum objects of which it is commonly said to be composed. It follows that neither quantum objects nor measuring instruments can be regarded as property-carriers existing independently of conscious subjects. However, if the difference between the classical domain and the quantum domain is understood as essentially the difference between the manifested world and what is instrumental in its manifestation, the elision of the subject can again be achieved.

\end{abstract}

\section{Introduction}
While the injection of quantum mechanics into a direct realism such as that defended by Searle \citeyearpar{Searle} leaves one with no choice but to espouse Bohr's concept of contextuality (Sec.~\ref{sec_Searle}), it presents one with the following choice: either one regards the experimental context in which a quantum phenomenon occurs as belonging to the furniture of a self-existent world, or one acknowledges, as Bohr did \citep{BohrOIM}, that the experimental arrangement owes much of its ``being thus'' \citep{Einstein-Dialectica} to the spatiotemporal structure of human sensory experience and to the logical or grammatical structure of human thought or language. 

The first option confronts one with the nonlocal features of quantum objects and their trans-phenomenal identity, but it offers no scope for accounting for them, besides insisting that ``the click in a counter is a totally lawless event, which comes by itself'' \citep{Ulfbohr}, or stressing that ``[o]bservations are to be regarded as discrete, disconnected events'' \citep{SchrSH} (Sec.~\ref{sec_first}). 

The second option makes it possible to address issues beyond the purview of direct realism with regard to experimental arrangements. In particular, it makes room for Bohr's insight that that the spatio\-temporal resolving power of human sensory experience is \textit{physically} limited by Planck’s constant, and that the concepts at our disposal, depending as they do on the spatiotemporal structure of human sensory experience, therefore can only be employed as far as sensory experience can reach (Sec.~\ref{sec_second}).

In recent papers \citep{BohrOIM,QBO,B4B} I argued that, owing to the contextuality of the properties of quantum objects, quantum mechanics does not countenance the elision of the thinking and perceiving subject, i.e., the possibility of ``stepping back into the role of a non-concerned observer'' \citep{SchrPO}. If quantum objects owe their properties to the experimental conditions in which they are observed, the experimental apparatus cannot owe its properties to the quantum objects of which it is commonly said to be composed. It follows that neither atoms and subatomic particles nor measuring instruments can be regarded as property-carriers that exist independently of conscious subjects. 

In a series of (more or less) recent papers \citep{MQW,QMMW,QMNL,QBO}, I also argued that the difference between the classical domain (containing objects which are directly accessible to sensory experience) and the quantum domain (containing objects which are at best indirectly accessible to sensory experience) is essentially the difference between the manifested world and what is instrumental in its manifestation. What I so far failed to state clearly was the intimate connection between these two lines of reasoning: if we envision the world’s manifestation as proceeding from a single ontological substance, then both quantum objects and measuring instruments owe their existence to this substance, in which case the elision of the subject can again be achieved (Sec.~\ref{sec_manifest}).

Sections \ref{sec_shapes} and \ref{sec_theone} sharpen the argument for the aforementioned interpretation of the difference between the classical domain and the quantum domain (relative to previous efforts), while Sec.~\ref{sec_objections} responds to some possible objections.

\section{Searle's table -- Eddington's tables -- contextuality (of the Bohrian kind)}\label{sec_Searle}
Searle \citeyearpar{Searle} begins by invoking the fact that we are able to communicate with each other using publicly available meanings in a public language. For this to work, we have to assume the existence of publicly available objects of reference:

\bq So, for example, when I use the expression ``this table'' I have to assume that you understand the expression in the same way that I intend it. I have to assume we are both referring to the same table, and when you understand me in my utterance of ``this table'' you take it as referring to the same object you refer to in this context in your utterance of ``this table.'' 

\eq The implication then is that

\bq you and I share a perceptual access to one and the same object. And that is just another way of saying that I have to presuppose that you and I are both seeing or otherwise perceiving the same public object. But that public availability of that public world is precisely the direct realism that I am here attempting to defend. (p. 276)

\eq As you and I are able to share a perceptual access to one and the same table, so you and I are able to share a perceptual access to one and the same measurement apparatus and thus, more specifically, to one and the same indicative event. An \textit{indicative event} is any event that indicates or is capable of indicating (making knowable or known) either the possession of a property by a quantum system or the possession of a value by a quantum observable. We can objectify our respective experiences of these things; we can think of the apparatus and the indicative event as part of the furniture of an objective world; we can pretend that they are what we perceive them to be \textit{independently} of the logical structure of human thought or the grammatical structure of our language, and \textit{independently} of the spatiotemporal structure of human sensory experience.

When quantum mechanics came along, the possibility of such a pretense ceased to exist. In a way it ceased to exist earlier, with the scientific ``revelation'' that things are not as they seem. Thus Eddington \citeyearpar{Eddington} distinguished between two tables, one that has extension, is comparatively permanent, colored, and substantial, and another ``scientific'' one which is mostly emptiness: ``Sparsely scattered in that emptiness are numerous electric charges rushing about with great speed.''

The difference between Eddington’s tables, however, pales in comparison to the difference between a quantum system and a measurement apparatus, inasmuch as the properties of quantum systems are \textit{contextual}. They are defined in terms of the experimental arrangements by which their presence is indicated, and they only exist actually and concretely (rather than potentially and abstractly) if and when their existence is indicated. The same cannot be said of Eddington’s tables, nor even of the electric charges he imagined rushing about with great speed.

Hitching the definition of the properties of quantum systems to the experimental conditions in which they are observed is Bohr’s ground-breaking contribution to science. It has, moreover, been spectacularly borne out by numerous no-go theorems including those of Bell \citeyearpar{Bell1,Bell2}, Kochen and Specker \citeyearpar{KS}, Greenberger et al. \citeyearpar{GHZ}, and Klyachko and coworkers \citeyearpar{KLYA}.

Not only the possessed properties of quantum systems are contextual. Atoms and subatomic particles in particular are contextual in that they are \textit{individuated} by the experimental contexts in which they are observed. Nobody has made this point more cogently than Falkenburg in her monograph \textit{Particle Metaphysics} [8], in which she concludes that ``only the experimental context (and our ways of conceiving of it in classical terms) makes it possible to talk in a sloppy way of quantum objects.'' While quantum objects

\bq seem to be Lockean empirical substances, that is, collections of empirical properties which constantly go together\dots, they are only individuated by the experimental apparatus in which they are measured or the concrete quantum phenomenon to which they belong\dots. Without a given experimental context, the reference of quantum concepts goes astray. In this point, Bohr is absolutely right up to the present day. (pp.~205--6)

\eq According to Falkenburg (p.~200),

\bq [q]uantum concepts refer to the properties of quantum phenomena which occur by means of a given experimental setup in a given physical context\dots. The carrier of these properties is nothing but the quantum phenomenon itself, say, a particle track on a bubble chamber photograph or the interference pattern of the double slit experiment.

\eq
\section{First option -- Genuine fortuitousness -- particles as instantaneous events}\label{sec_first}
Now we have a choice. We may speak of the experimental context in which a quantum phenomenon occurs in the same way as Searle speaks of a table, to wit, as belonging to the furniture of a self-existent world. Or we may acknowledge that tables and experimental contexts owe much of their \textit{So-Sein}---their ``being thus'' \citep{Einstein-Dialectica}---to the spatiotemporal structure of human sensory experience and to the logical/grammatical structure of human thought/language.

The first option confronts us with the nonlocal features of quantum objects and their trans-phenomenal identity which, not only to Falkenburg (p. 206), ``remain mysterious.'' An efficient way of disposing of the nonlocal features has been proposed by Ulfbeck and Bohr \citeyearpar{Ulfbohr}. To the question ``Where does that click come from?'' they gave the answer: ``it comes by itself, out of the blue.'' Also, it is ``an event entirely beyond law.'' (What is not beyond law is the statistical correlations between clicks.) While clicks are ``events in spacetime,'' there are no particles ``on the spacetime scene.'' Genuinely fortuitous events, occurring by themselves, form ``the basic material that quantum mechanics deals with.''

It is worth noting that \textit{every} successful measurement in quantum mechanics is genuinely fortuitous, not just with regard to its outcome but with regard to its very occurrence. While quantum-mechanical probability assignments are made on the assumption that there will be an outcome (failing which there will be no measurement), nothing guarantees the success of an attempted measurement. The redundancy typically built into measuring devices can maximize the likelihood of success, but it cannot guarantee success.

Ulfbeck and Bohr also dispose of the trans-phenomenal identity of quantum objects, to which Misner, Thorne, and Wheeler \citeyearpar[p. 1215]{MTW} refer as ``the miraculous identity of particles of the same type.'' Ulfbeck and Bohr cut this Gordian knot by concluding that ``clicks can be classified as electron clicks, neutron clicks, etc.'' even though ``there are no electrons and neutrons on the spacetime scene.'' Schr\"odinger \citeyearpar{SchrSH} made much the same point, stressing that the idea that a particle is an enduring entity must be dismissed:

\bq We are now obliged to assert that the ultimate constituents of matter have no ``sameness'' at all. When you observe a particle of a certain type, say an electron, now and here, this is to be regarded in principle as an \textit{isolated event}. Even if you do observe a similar particle a very short time later at a spot very near to the first \dots there is no true, unambiguous meaning in the assertion that it is \textit{the same} particle you have observed in the two cases. (p. 121, original emphases)

\medskip Observations are to be regarded as discrete, disconnected events. Between them there are gaps which we cannot fill in. There are cases where we should upset everything if we admitted the possibility of continuous observation. That is why I said it is better to regard a particle not as a permanent entity but as an instantaneous event. Sometimes these events form chains that give the illusion of permanent beings---but only in particular circumstances and only for an extremely short period of time in every single case. (pp. 131--32)

\eq
\section{Second option -- the metaphysical import of Planck's constant -- limits to the objectivation of  spatial distinctions}\label{sec_second}
If we chose the second option, acknowledging that tables and experimental arrangements owe much of their \textit{So-Sein} to the spatiotemporal structure of human sensory experience and the logical or grammatical structure of human thought or language, we face another mystery, made famous by Einstein \citeyearpar[p. 292]{EinsteinIO}:

\bq The very fact that the totality of our sense experiences is such that by means of thinking (operations with concepts, and the creation and use of definite functional relations between them, and the coordination of sense experiences to these concepts) it can be put in order, this fact is one which leaves us in awe, but which we shall never understand. One may say ``the eternal mystery of the world is its comprehensibility.'' It is one of the great realisations of Immanuel Kant that the setting up of a real external world would be senseless without this comprehensibility.%
\footnote{This is from a translation of an article Einstein \citeyearpar{EinsteinPR} wrote in German, in which the penultimate sentence reads: ``Man kann sagen: Das ewig Unbegreifliche an der Welt ist ihre Begreiflichkeit.'' The rhetorically strong juxtaposition of ``Das ewig Unbegreifliche an der Welt'' (the eternally incomprehensible thing about the world) and ``ist ihre Begreiflichkeit'' (is its comprehensibility) has been lost in this translation. The popular internet version---``The most incomprehensible thing about the world is that it is comprehensible''---is actually more faithful to the original than the ``official'' translation by Jean Piccard.}

\eq Kant's insight that what we call ``the world'' is a construct capable of being objectified and shared, rather than something that exists in itself, independently of thinking and perceiving subjects, makes it possible to address issues beyond the purview of direct realism with regard to objects to which we have perceptual access. It allowed Kant to position freedom and moral responsibility alongside the causality requisite to comprehending the world. And it allowed Bohr to grasp the metaphysical import of Planck's constant.

Kant’s theory of science did not allow for a limit to the spatiotemporal resolving power of human sensory experience. One could conceive of the empirical world as partitioned into infinitesimal regions of space and infinitesimal intervals of time. Physical objects could be taken to possess exact positions, and physical events could be taken to happen at exact times, as they were by Newton and classical physics in general. One essential departure of Niels Bohr’s thinking from Kant’s, which engenders most of the others, is his insight that the spatiotemporal resolving power of human sensory experience is \textit{physically} limited by Planck’s constant.

If sensory experience could in principle reach all the way down to infinitesimal regions of spacetime, we could detach ourselves from our shared external world in the manner first conceived by Kant. Having asserted that ``we can have cognition of no object as a thing in itself, but only insofar as it is an object of sensible intuition, i.e. as an appearance,'' Kant \citeyearpar[p.~115]{Kant} could go on to affirm, sensibly enough, that ``even if we cannot cognize these same objects as things in themselves, we at least must be able to think them as things in themselves. For otherwise there would follow the absurd proposition that there is an appearance without anything that appears.''

If the spatiotemporal resolution of sensory experience is limited, and if the construction of the objective world depends on concepts that owe their meanings in part to the spatiotemporal structure of human sensory experience, then these concepts can only be employed, and the construction of a shared objective world can only be carried, as far as sensory experience can reach. The only meaningful way of talking about quantum objects is then in terms of correlations between indicative events. In the words of Bohr: ``the physical content of quantum mechanics is exhausted by its power to formulate statistical laws governing observations obtained under conditions specified in plain language'' \citeyearpar[p.~159]{Bohr10};
``the quantum-mechanical formalism \dots\ represents a purely symbolic scheme permitting only predictions \dots\ as to results obtainable under conditions specified by means of classical concepts'' \citeyearpar[pp. 350--51]{Bohr7}.%
\footnote{To Bohr, classical concepts are classical not because they are proprietary to classical physics but because we know what they mean, inasmuch as their meanings are rooted in what we all have in common, to wit, the spatiotemporal structure of sensory experience and the grammatical structure of plain, ordinary language. \citep{BohrOIM}}

\section{Manifestation -- particles -- a single ontological substance -- elision of the subject}\label{sec_manifest}
While Bohr did stop there, we are under no obligation to follow suit. This does not mean that we should try to construct macroscopic reality from the bottom up---a futile strategy, as Falkenburg has shown. In a series of (more or less) recent papers \citep{MQW,QMMW,QMNL,QBO}, I argued that the difference between the classical domain (containing objects which are directly accessible to sensory experience) and the quantum domain (containing ``objects'' which are at best indirectly accessible to sensory experience) should be interpreted as the difference between the manifested world and what is instrumental in its manifestation. I envisioned the (atemporal) process of manifestation as a progression from the undifferentiated unity of \textit{a single ontological substance} to a world which allows itself to be described in the plain language of everyday discourse. Subatomic particles, non-visualizable atoms, and partly visualizable molecules, instead of being constituent parts of the manifested world, mark the stages of this progression.

How, in this framework, can the intermediate stages of that progression be described? The first thing to note here is that the question does not concern \textit{individual} quantum objects; it concerns object \textit{types}. The second is that the descriptions can only be given in terms of correlations between indicative events. And the third is that these correlations can only be used counterfactually. 

For a basic example, consider a stationary ``state'' of atomic hydrogen. (The scare quotes serve as a reminder that quantum-mechanical states are conditional probability algorithms, not states in the classical sense.) Such a state is conditional on the outcomes of three measurements---the atom's energy, its total angular momentum, and a component of its angular momentum (whose definition involves the orientation of a measurement apparatus)---and  it serves to assign probabilities to the outcomes of possible measurements, such as a measurement of the electron's position relative to the nucleus. 

The long and the short of it is that what is instrumental in the manifestation of the empirical world can only be described in terms of statistical correlations between events that happen (or could happen) in the empirical world. This goes a long way towards explaining why the general theoretical framework of contemporary physics is a probability calculus, and why the events to which it serves to assign probabilities are indicative events. What else does it buy us?

I used to argue that, owing to the contextuality of the properties of quantum objects, quantum mechanics does not countenance the elision of the thinking and perceiving subject, not even to the extent Kant’s transcendental philosophy of science did  \citep{BohrOIM,QBO,B4B}. Here is how Schr\"odinger \citeyearpar{SchrPO} expressed this possibility:

\bq Without being aware of it and without being rigorously systematic about it, we exclude the Subject of Cognizance from the domain of nature that we endeavour to understand. We step with our own person back into the part of an onlooker who does not belong to the world, which by this very procedure becomes an objective world.

\eq If quantum objects owe their properties to the experimental conditions in which they are observed, the experimental apparatus cannot owe its properties to the quantum objects of which it is commonly said to be composed. It follows that neither atoms and subatomic particles nor measuring instruments can be regarded as property-carriers that exist independently of conscious subjects. The existence of both kinds of objects will have to be attributed to subjects which perceive and/or mentally construct them. But if we envision the world’s manifestation as proceeding from a single ontological substance, then both quantum objects and measuring instruments owe their existence to this substance. And in this case the elision of the subject can again be achieved, at least to the extent Kant did achieve it.

\section{The shapes of things -- formless particles -- the miraculous identity of particles of the same type -- adding probabilities or adding amplitudes}\label{sec_shapes}
But how can the claim that the manifestation of the empirical world proceeds from a single ontological substance---let's call it ``the One''---be justified? What needs to be demonstrated is that the manifested world, as well the things it contains, owe their spatial multiplicity to spatial relations that obtain between the One and the One: by entering into \textit{reflexive} spatial relations, the One gives rise to (i)~what looks like a multiplicity of relata if the reflexive quality of the relations is ignored, and (ii)~what looks like a substantial expanse if the spatial quality of the relations is reified. In other words, if by ``space'' we mean the totality of existing spatial relations, and if by ``matter'' we mean the corresponding (apparent) multitude of relata, then by entertaining reflexive spatial relations the One founds both matter and space.

How can we demonstrate this? To begin with, the fundamental particles of the Standard Model are often described as pointlike, and while some appear to take this to be the literal truth, it can only mean that they are \textit{formless}. To see this, recall that one essential departure of Bohr’s thinking from Kant’s is his insight that the spatiotemporal resolving power of human sensory experience is \textit{physically} limited by Planck’s constant. Accordingly, there are limits to the extent to which physical space can be objectively (i.e., objectifiably) partitioned. But the existence of a literally pointlike object would imply that physical space is intrinsically partitioned ``all the way down.'' There would be a location at which such an object is situated, and there would be another location infinitesimally close to it at which nothing is situated.

The shapes of things thus resolve themselves into sets of spatial relations that obtain between formless relata. This takes us back to ``the miraculous identity of particles of the same type,'' which Misner, Thorne, and Wheeler regarded as ``a central mystery of physics.'' To illustrate the issue, suppose that initially we observe a particle of type $X$ at location $A$ and a particle of type $Y$ at location $B$, and that subsequently we observe a particle of type $X$ at location $C$ and a particle of type $Y$ at location $D$. The mystery is that if the particles observed are of the same type (i.e., $X=Y$), then it is demonstrably false (i.e., inconsistent with the quantum statistics) to assume that the particle found at $A$ is numerically identical with either the particle found at $C$ or the particle found at~$D$ (the way the Morning Star is numerically identical with the Evening Star). 

While this alleged mystery is easily demystified, for it only exists if we wrongly assume that particles are intrinsically distinct and endure, there is, underlying it, another mystery. If $X=Y$, we  calculate the probability of finding a particle at $C$ and a particle at $D$, given one was found at $A$ and one at $B$, by first adding two amplitudes and then taking the absolute square of the result. If instead \hbox{$X\neq Y$}, we calculate the same probability by first taking the absolute squares of the same amplitudes and then adding the results. Whence this difference?

To find out, we first need to recapitulate the steps that take us from the classical probability calculus (in its simplest form, a point in a phase space) to the quantum-mechanical one (in its simplest form, a 1-dimensional subspace in a complex vector space). Having defined ``ordinary'' material objects as objects that (i)~have spatial extent (they ``occupy space''), (ii)~are composed of a (large but) finite number of objects that lack spatial extent (particles that do not ``occupy space''), and (iii)~are stable (they neither explode nor collapse as soon as they are created), we explored what the existence of such objects implies \citep[Chapter 7]{Mohrhoff-book2}. 

The first implication is that the product of the standard deviations of relative positions and their corresponding momenta must have a positive lower limit.%
\footnote{See also \cite[Section 25.1]{Mohrhoff-book2}.}
This calls for nontrivial probabilities (i.e., probabilities greater than $0$ and less than $1$). As the classical probability calculus cannot accommodate probabilities that are both irreducible and nontrivial, we need to find a probability calculus that can, and the most straightforward way to make room for nontrivial probabilities is to upgrade from a 0-dimensional point in a phase space to a 1-dimensional subspace in a complex vector space.%
\footnote{The virtual inevitability of this upgrade has been demonstrated by Jauch \citeyearpar[pp.\ 92--94, 132]{Jauch}.}
Three conclusions ensue, which together are sufficient to prove the trace rule, of which the Born rule is a special case. The first conclusion is that measurement outcomes are represented by the projectors of a complex vector space~$\cal V$. The second is that the outcomes of compatible elementary measurements correspond to commuting projectors. And the third is that if $\mathbf{P}_1$ and~$\mathbf{P}_2$ are orthogonal projectors, then the probability of the outcome represented by $\mathbf{P}_1+\mathbf{P}_2$ is the sum of the  probabilities of the outcomes represented by $\mathbf{P}_1$ and~$\mathbf{P}_2$. 

Suppose, then, that a measurement performed at a given time yields the outcome $u$ represented by $\mathbf{P}_1=|u\rangle\langle u|$, and that we want to calculate the probability with which a measurement performed at a later time yields the outcome $w$ represented by $\mathbf{P}_2=|w\rangle\langle w|$. Further suppose that at some intermediate time another measurement with possible outcomes $v_i$ is made. If the outcome of the intermediate measurement is ignored, then by the Born rule and the law of total probability, the wanted probability is given by 
\begin{equation}
	p(w|u)=\sum_i|\langle w|v_i\rangle \langle v_i|u\rangle|^2.
\end{equation}
In words: we first calculate the absolute squares of the amplitudes $A_i=\langle w|v_i\rangle \langle v_i|u\rangle$ and then add the results. If the intermediate measurement is not made, we instead have that
\begin{equation}
p(w|u)=|\langle w|u\rangle|^2=\Bigl|\sum_i \langle w|v_i\rangle \langle v_i|u\rangle\Bigr|^2.
\end{equation}
In words: we first add the amplitudes $A_i$ and then take the absolute square of the result. 

The difference between the two rules for calculating $p(w|u)$ thus is a feature of a theoretical framework whose validity is implied by the mere existence of ``ordinary objects.'' In the case of a pair of particles detected at $A$ and $B$ and a pair of particles subsequently detected at $C$ and~$D$, $X$~and $Y$ (if different) play the role of the intermediate measurement. They create an objective difference between two alternatives, one where $X$ and $Y$ are instantiated at $C$ and~$D$, respectively, and one where they are instantiated at $D$ and~$C$, respectively. $X$~and $Y$ are often interpreted as ``identity tags,'' an error I have committed myself more than once. If for the two amplitudes we write $a(C{=}A, D{=}B)$ and $a(C{=}B, D{=}A)$, care must be taken to read the equals sign as expressing not numerical identity but merely falling under the same type.

\section{Some possible objections}\label{sec_objections}
A couple of issues might be raised against these claims. In non-relativistic quantum mechanics, the probability of finding an isolated particle is conserved. The non-relativistic theory, however, is a hybrid of of classical and quantum concepts; to address the issue of the identity of successively observed particles, we need to consult the relativistic theory.

In relativistic particle physics, what is conserved is not property carriers but dynamical properties including energy-momentum and several types of charge. Relativity requires conservation laws to be local, and quantum field theory (which deals with particle types and their interconversions) adds to this that the conserved dynamical properties come bundled in specific ways. The bundles or types, however, show up as individuals or tokens only in specific experimental contexts. In a typical scattering experiment, a set of incoming particles transforms into a set of outgoing particles in such a way that the conserved quantities (i)~are separately conserved and (ii)~only appear in type-specific combinations. The $S$-matrix, used to calculate the transition probability from a given set of incoming particles to a given set of outgoing particles, treats the scattering itself as a black box. In the words of Falkenburg \citeyearpar[pp. 131--32]{Falkenburg}: 

\bq What is going on inside the black box, i.e., during the scattering, is illustrated in terms of Feynman diagrams. However, they have no literal meaning. They are mere iconic representations of the perturbation expansion of a quantum field theory. They make the calculations easier, but they do not represent individual physical processes.

\eq The perturbation expansion is a calculational tool. It sums over amplitudes, each ``telling a story'' how the incoming particles collide, create, and/or annihilate particles in such a way that a specified set of outgoing particles is produced. None of these stories has anything to do with what (if anything) actually goes on inside the box. It follows that the type-identifiers $X$ and $Y$attached to the particles detected at $A$, $B$, $C$, and $D$ cannot be interpreted as token-identifiers. 

Another issue that might be raised concerns the identification of clicks. If ``clicks can be classified as electron clicks, neutron clicks, etc.'' even though ``there are no electrons and neutrons on the spacetime scene,'' as Ulfbeck and Bohr have stressed, we should be able to know an electron click when we see (or hear) one. Yet a single click does not usually announce the type to which it belongs. Its type has to be inferred from a sequence of detection events, such as an optically recorded sequence forming a track in a bubble, cloud, spark, or streamer chamber. A sequence of detection events makes it possible to measure such quantities as the radius of curvature of a particle's track in a magnetic field, a particle's time of flight, a particle's kinetic energy, or a particle's energy loss through ionization and excitation. Measuring three of these four quantities is in principle sufficient to positively identify the type of particle forming the track \cite[Chapter~9]{GS2008}, which then makes it possible to classify the individual detection events.

The reference of ``particle'' in the previous paragraph, however, is ambiguous. The attribution of physical properties to subatomic particles is not based on a single unified theory but on several incommensurable theories: ``the operational, axiomatic, and referential aspects of physical concepts fall apart'' \cite[p.~162]{Falkenburg}. The axiomatic (i.e., field and group theoretical) concepts only concern particle types. The operational concepts pertain to the methods by which predictions derived from the axioms are tested, but they agree with the axiomatic concepts only at the probabilistic level. A bridge principle is needed to align them with the referential concepts, which apply to individual particles via quantum phenomena which occur in experimental contexts. But the only bridge principle with referential import is the correspondence principle (p.~198). It  makes it possible to express quantum phenomena in terms of classical physical quantities (p.~206). In particular, the quantum-mechanical particle track corresponds to a classical path:

\bq To be more precise, the quantum mechanics of scattering predicts an \textit{ensemble of possible tracks} which are \textit{identical for all practical purposes} and which correspond to a classical trajectory\dots.  The probability of two subsequent particle deflections along a straight line differs from 1 by an extremely small value which may be neglected \textit{for all practical purposes}. (pp. 176--77)

\medskip Only the correspondence principle claims that the quantum mechanics of scattering refers to individual particle tracks and subatomic scattering centers. Where correspondence breaks down, that is, in the relativistic domain, the physical interpretation of individual particle tracks and scattering events is only supported by symmetries, conservation laws, and superselection rules. But all of these principles work only for the particle \textit{type}. They do not refer at the token level, that is, to \textit{individual} particles and measurement results. (p.~198)

\medskip Quantum theory in general (including the current relativistic quantum field theories of high energy physics) has \textit{no realistic interpretation} which establishes reference to individual systems. Strictly speaking, up to the present day \textit{there is no quantum theory of individual systems}. Only by means of superselection rules, the generalized correspondence principle, and so on, does quantum physics apply to individual systems. (p.~207)

\eq Thus in the non-relativistic regime individual particles evince themselves---\textit{for all practical purposes}%
\footnote{In Chapter 5 of Falkenburg \citeyearpar{Falkenburg}, titled ``Measurement and the Unity of Physics,'' the expression ``for all practical purposes'' occurs eleven times.}---%
as tracks, from which the types to which they belong can be inferred. In the relativistic regime, they only evince themselves as clicks, between which ``there are gaps we cannot fill in''  \cite[p.~131]{SchrSH}.

\section{The One}\label{sec_theone}
Consider once more the two pairs of particles---one found at $A$ and $B$ in possession of $X $and~$Y$, respectively, the other subsequently found at $C$ and $D$ also in possession of $X$ and~$Y$, respectively. If we say that what is observed first is \textit{two} things with \textit{two} pairs of properties (being at $A$ as an $X$ and being at $B$ as a~$Y$), then we use the word ``two'' once too often. The question is not merely: which of the first observed particles is identical with which of the subsequently observed particles? The question already arises for a single pair; it concerns whether two simultaneously observed particles---regardless of the types to which they belong---are \textit{two things}. And the answer is that since quantum ``objects'' are neither self-existent substances nor property carriers, nothing warrants the claim that, in addition to the click of type $X$ at $A$ and the click of type $Y$ at~$B$, there are two \textit{things}.

In Sec.~\ref{sec_manifest}~ we came to the conclusion that the elision of the subject can once again be achieved, \textit{provided} that we think of the manifestation of the world as an (atemporal) process that begins with a single ontological substance (``the One''). To demonstrate the cogency of this idea, one has to show that the manifested world (as well as each of the things it contains) owes its spatial multiplicity, not to a multitude of things, but to a multitude of self-relations---relations that obtain between the One and the One. In Sec.~\ref{sec_shapes}~ we came to the conclusion that the shapes of things resolve themselves into sets of spatial relations obtaining between formless relata. What remains to be shown is that the spatial relations obtaining between these formless relata are reflexive, or that in reality there is but a \textit{single} formless relatum, or that the relations obtain between relata that are not only formless but also \textit{numerically identical}.

Now, there are \textit{two} ways of explaining why, if $X{=}Y$, there is no answer to the question: which of the initially observed particles is identical with which of the subsequently observed ones? The first is to argue that \textit{no two} particles are ever identically the same thing, as we just did. The second is to argue that \textit{all} particles are identically the same thing. What we cannot do is pick and choose; if~$X{=}Y$, we cannot say that the particle at $A$ is (numerically) identical \textit{only} with the particle at~$C$.

If we want to think of the manifestation of the world as an (atemporal) process that begins with a single ontological substance, we must choose the second option. As far as the two pairs of particles are concerned, there is but one thing, which is initially observed both at~$A$ (as an~$X$) and at~$B$ (as a~$Y$), and which is subsequently observed both at~$C$ (as an~$X$) and at~$D$ (as a~$Y$). As far as the manifestation of the world is concerned, there is but a single, indivisible, formless substrate, whose reflexive spatial relations constitute the shapes of things. What is more, this single ontological substance can also be thought of as the sole agent that causes every click, and that in fact is responsible for the occurrence of every measurement. (As noted in Sec.~\ref{sec_first}~, the actual occurrence of a quantum-mechanical measurement would otherwise remain unexplained.)

\section{Summary and Conclusion}
In several recent papers I argued that, owing to the contextuality of the properties of quantum objects, quantum mechanics does not countenance the elision of the thinking and perceiving subject: if quantum objects owe their properties to the experimental conditions in which they are observed, the experimental apparatus cannot owe its properties to the quantum objects of which it is commonly said to be composed, whence it follows that neither quantum objects nor measuring instruments can be regarded as property-carriers that exist independently of conscious subjects. 

In a series of (more or less) recent papers I also argued that the difference between the classical domain  and the quantum domain is essentially the difference between the manifested world and what is instrumental in its manifestation. What I so far failed to state clearly was the intimate connection between these two lines of reasoning: if we envision the world's manifestation as proceeding from a single ontological substance, then both quantum objects and measuring instruments owe their existence to this substance, in which case the elision of the subject can again be achieved. This has been the central point of the present paper. In addition, the argument for this interpretation of the difference between the classical and quantum domains (relative to previous efforts) has been sharpened, and some possible objections have been met.

\raggedright

\end{document}